# Electrical Characterization of CIGS Thin Film Solar Cells by Two and Four-Wires Probe Technique

Amirhosein Mosavi [1], Beszedes Bertalan [2], Felde Imre [1], Laszlo Nadai [1], Nima E. Gorji [3]*

[1] *Kalman Kando Faculty of Electrical Engineering, Obuda University, Budapest, Hungary; amir.mosavi@kvk.uni-obuda.hu*

[2] *Alba Regia Technical Faculty, Obuda University, Szekesfehervar, Hungary; beszedes.bertalan@amk.uni-obuda.hu*

[3] *Optoelectronics Research Group, Faculty of Electrical and Electronics Engineering, Ton Duc Thang University, Ho Chi Minh City, Viet Nam*

*\*Correspondence: nimaegorji@tdtu.edu.vn*

**Abstract**

The characterization of thin-film solar cells is of huge importance for obtaining high open-circuit voltage and low recombination rates from the interfaces or within the bulk of the main materials. Among the many electrical characterization techniques, the two- and four-wire probe using the Cascade instrument is of interest since the resistance of the wires, and the electrical contacts can be excluded by the additional two wires in 4-wire probe configuration. In this paper, both two and four-point probes configuration are employed to characterize the CIGS chalcogenide thin-film solar cells. The two-wire probe has been used to measure the current-voltage characteristics of the cell which results in a huge internal resistance. Therefore, the four-wire connection are also used to eliminate the lead resistance to enhance the characterization's accuracy. The load resistance in the two-wire probe diminishes the photogenerated current density at smaller voltage ranges. In contrast, the proposed four-wire probe collects more current at higher voltages due to enhanced carrier collection efficiency from contact electrodes. The current conduction mechanism is also identified at every voltage region represented by the value of the ideality factor of that voltage region. It is observed that a longer time given to the charge collection results in increased current density at a higher voltage. According to the results and device characteristics, a novel double-diode model is suggested to extract the saturation current density, shunt and series resistances and the ideality factor of the cells. These cells are shown to be efficient in terms of low recombination at the interfaces and with lower series resistance as the quality of the materials is in its most possible conductive form. The measured internal resistance and saturation current density and ideality factor of the two measurement configuration are measured and compared.

## I.     Introduction

Copper indium gallium selenide (CIGS) thin-films solar cells are made of the compounds of chalcopyrite-type ternary, i.e., $CuInS_2$, $CuInSe_2$ (CIS) and $CuGaSe_2$ (CGS), with the bandgaps of 1.5 eV, 1.0 eV and 1.7 eV, respectively [1]. The idea then came up to make a multi-alloy layer of Cu (In, Ga)(Se, S)$_2$ belongs to the I-III-VI2 family, which together forms a CIGS absorber layer with a tunable bandgap of around 1.3-1.55 eV depending on the composition of Ga or In [2,3]. The energy conversion efficiency of over 23% has been reported for a lab-scale CIGS cell by the Chinese Academy of Sciences (ISCAS) [4,5]. Nevertheless, the performance of industrial-scale or modules of the same materials is between 15% to 18%. The characteristics of the bandgaps and mesh parameters can be modified through replacing the atoms of either indium or selenium with, e.g., S for Se, and Ga for In. Thus, there exist bandgap energies with grades of 1.04 eV for $CuInSe_2$ and 2.4 eV for $CuGaS_2$. The most popular CIGS layer deposition techniques are concerned with the reactive cathodic co-sputtering or the co-evaporation of the extremely heated Cu-In-GaSe-S elements. Such techniques empower high performance and efficient control over the CIGS layers' crystal growth [6]. Alternatively, the low-temperature deposition accompanied by sulfurization, selenization, and subsequent annealing can also be used for inkjet printing, sputtering, and electrodeposit. The latter methods need post-growth annealing to achieve a high-quality absorber layer with acceptable optoelectronic characteristics. The CIGS cell efficiency decreases by recombination to midgap defect states or at dangling bonds. Also, the band discontinuities at CIGS interfaces cause inefficient carrier collection. Na incorporation improves $V_{oc}$, FF, efficiency, by improving the CIGS film morphology and grains [7]. It also improves the holes concentration/conductivity by creating acceptor type NaIn defect and passivates the grain boundaries which leads to higher Voc. The challenges with CIGS cells are as fowling: (i) to enhance the light absorption in CIGS layer (ie. by using innovative doping methods) to gain higher $J_{sc}$, (ii) improving the surface passivation on both interfaces of CIGS layer and reducing the defect density at the interfaces to minimize the interface recombination loss for a high Voc, (iii) using Cd-free large bandgap window layer for high $J_{sc}$, (iv) utilizing reflector (mirror) layer at the rear side of the cell to redirect unused light back to the absorber for electron-hole pair generation, (iv) Mo rear contact was reported to cause instability due to its oxidation and degradation thus alternative back contact materials are required to prevent oxygen and moisture ingression into the cell structure. The polycrystalline thin-film solar cells based on chalcopyrite Cu (In, Ga)$Se_2$ (CIGS) are widely under considerations from different aspects. We have recently considered theoretically some issues of such cells all of; bandgap grading both in conduction and valence bands [9], modeling of the recombination and generation mechanisms [10,11], Auger process [12]. Beside of such studies, the electrical properties of these devices can be presented in the form of equivalent circuits [13], where the electrical parameters can be obtained by admittance spectroscopy and electrical measurements.

In this contribution, we perform the current-voltage characterization of CIGS solar cells by two and four-point probe techniques available in the Cascade machine connected to a LabVIEW software that was programmed for a device metric extraction. The results are compared for the 2- and 4-wire probe measurements and show a time collection which is related to the internal device resistance.

## II.     Experimental

The solar cell devices in this study is a typical CIGS thin-film solar cell shown in Fig. 1 with the configuration of Glass/Mo/Cu (In, Ga)Se$_2$/CdS/ZnO/ZnO: Al. The structure and materials are shown in Table 1.

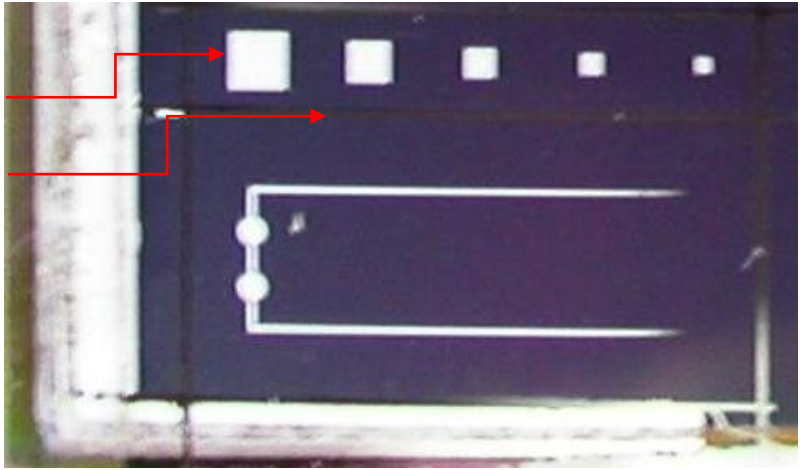

Figure 1, the CIGS thin-film solar cell understudy with the contacts shown with red arrows.

Table I. The device architecture and the thickness of the film layers.

| Glass | Mo | CIGS | CdS | ZnO | ZnO:Al |
|---|---|---|---|---|---|
| 5 mm | 2 μm | 150 nm | 80 nm | 100 nm | 180 nm |

A Cascade machine has been used for the JV characterization of the cell controlled by LabView package to draw the JV curves values for both 2- and 4-wire probe configurations. The Connection of 4200A-SCS to a solar cell for JV measurements is shown in Fig. 1 [14].

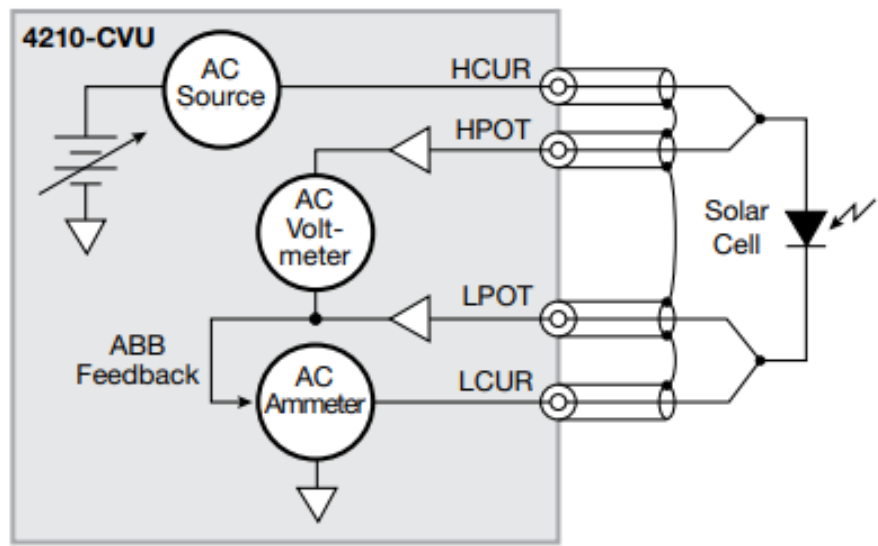

Figure 2, the 4-wire probe configuration used to characterize the cell.

We have placed the four tips of the probe electrodes on top and bottom contacts of the cell using the microscope handle and the switched light, as shown in Fig. 2. The cascade has translating gears to move the electrodes over the contact areas but must be careful not to scratch the cell surface or break out the cascade's electrode tips. The Cascade is connected to a PC, and a LabVIEW code has been developed to control and collect the current-voltage using a Source meter. We have measured the current-voltage of the cells by both 2- and 4-wire configuration dark and biased conditions. We switched off the light in order to measure the internal resistance of the cells without the interference of conductivity induction through photovoltage of light exposure. In the 4-wire measurement configuration, we applied the voltage to the top contacts (metallic contact) by 1 probe electrode while the fourth electrode tip is grounded. The current-voltage is collected by the other two probing electrodes applied on bottom contact (FTO). Note that we made a common for the two additional electrodes when measuring the two-wire configuration. Table 2 presents the current, voltage, fill factor, and performance of the cell measured by the two configuration. A conversion performance of 16.62% has been obtained from the cell which is reasonably high for such as architecture and demonstration.

Table 2. The device metrics of the CIGS thin-film solar cell under study.

| $\eta$ | FF | $J_{sc}$ | $V_{oc}$ |
|---|---|---|---|
| 16.61 % | 73.8 % | 35.04 mA/cm$^2$ | 605 mV |

### III. Results and Discussion

To measure the current-voltage characteristics of the CIGS thin-film solar cells are forward biased from 0 - 0.8 V at dark conditions and the collection time for every voltage step was kept at 0.1 s or 10 s in order to see the differences between the two probe configurations. We have measured the current-voltage characteristics of the cells at 300 K of room temperature by both 2 and 4-wire probe methods and the results are presented in Figs. 3-5.

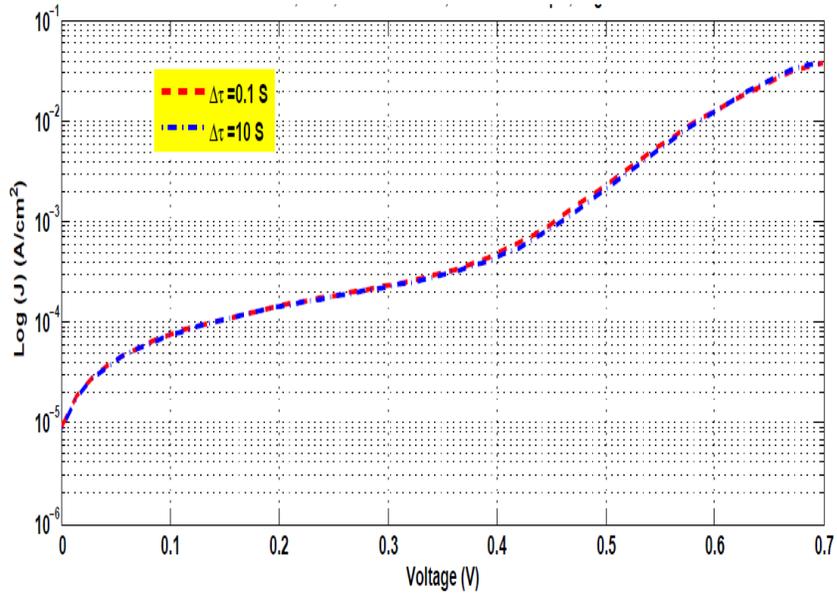

Figure 3, JV curve obtained with four-wire technique under Dark Forward biased.

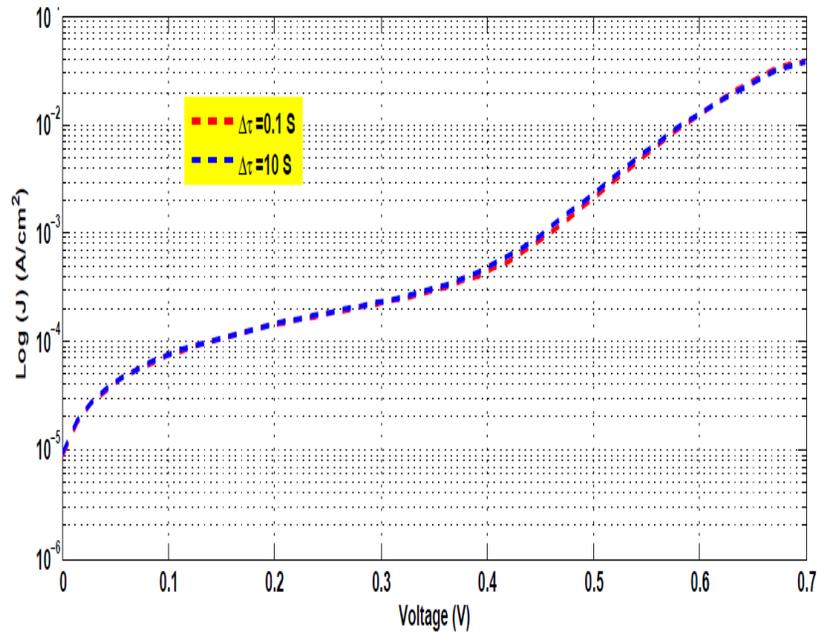

Figure 4, JV curve obtained with two-wire technique under Dark Forward biased.

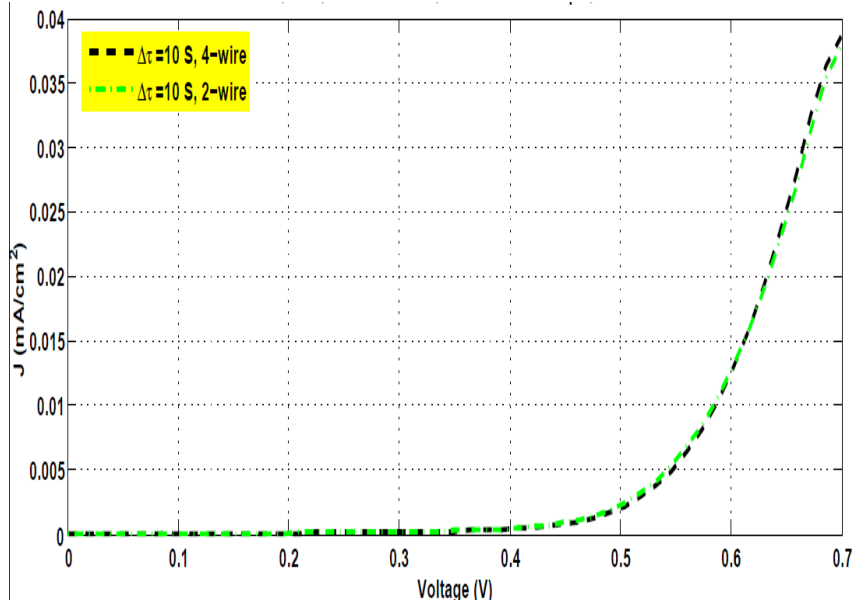

Figure 5, comparison of JV curves obtained by two and four-wire under dark or forward bias.

From the above presented figures, we can calculate the ideality factor for every voltage region. For example, for a voltage of 0.2 V, the JV equation is given by the following double-diode model as also shown in Fig. 6.

$$J = J_{01}\left\{\exp\left(\frac{q\vec{\tau}(V-IR_S)}{n_1 kT}\right) - 1\right\} + J_{02}\left\{\exp\left(\frac{q\vec{\tau}\vec{\tau}(V-IR_S)}{n_2 kT}\right) - 1\right\} + \frac{V-IR_S}{R_{sh}} \quad (1)$$

where $V$ is the applied voltage, $n$ is the diode ideality factor, $J_{o1,02}$ is the reverse saturation current density and $k$ is the Boltzmann's constant. Fig. 5 represents the double-diode model of the cell with $D_1$ and $D_2$ as diffusion, $R_s$ and $R_{sh}$ as series and shunt resistances.

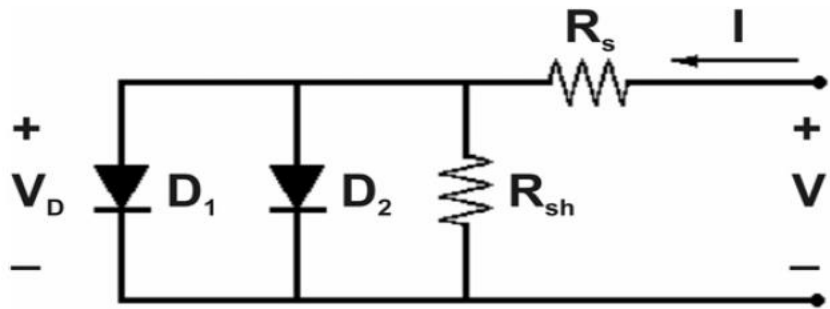

Figure 6, Double-diode model proposed for this CIGS thin-film solar cell.

The values of diode parameters such as $J_{01,02}$, $n$, $J_o$, $R_{sh}$, and $R_s$ are evaluated at the main voltage region are extracted in Table 3. We used the analytical approach presented by Das et al. in Ref. [15] to extract the electrical parameters of the model from these dark measurements.

Table 3. The Solar Cell Parameters extracted for the cell from our measurements.

| $J_{01}$ | $n_1$ | $J_{02}$ | $n_2$ | $R_{sh}$ | $R_s$ |
|---|---|---|---|---|---|
| $4.21\times10^{-6}$ mA/cm$^2$ | 1.68 | $3\times10^{-10}$ mA/cm$^2$ | 5.14 | 2.5 k$\Omega$cm$^2$ | 0.346 $\Omega$cm$^2$ |

The JV characteristic curves in the logarithmic format are presented in Figs. 3-5. The different voltage regions can be analyzed and be assigned to various conduction mechanisms as shown in Fig. 7. In every region, we may get a particular value for the ideality factor ($n$) and saturation current density ($J_0$). For every $n$ and $J_0$ we analyze the dominant current conduction mechanism. We have done this for other thin-film solar cells in our previous research outputs [16] or diffusion process. At room temperature (300K), a constant value for the ideality factor within $1 < n < 2$ is extracted which is attributed to interface recombination where the recombination current is dominating the charge transport. The reason for such an assignment is that this n value is related to the dielectric constant of the medium under study and the donor/acceptor concentration ratio.

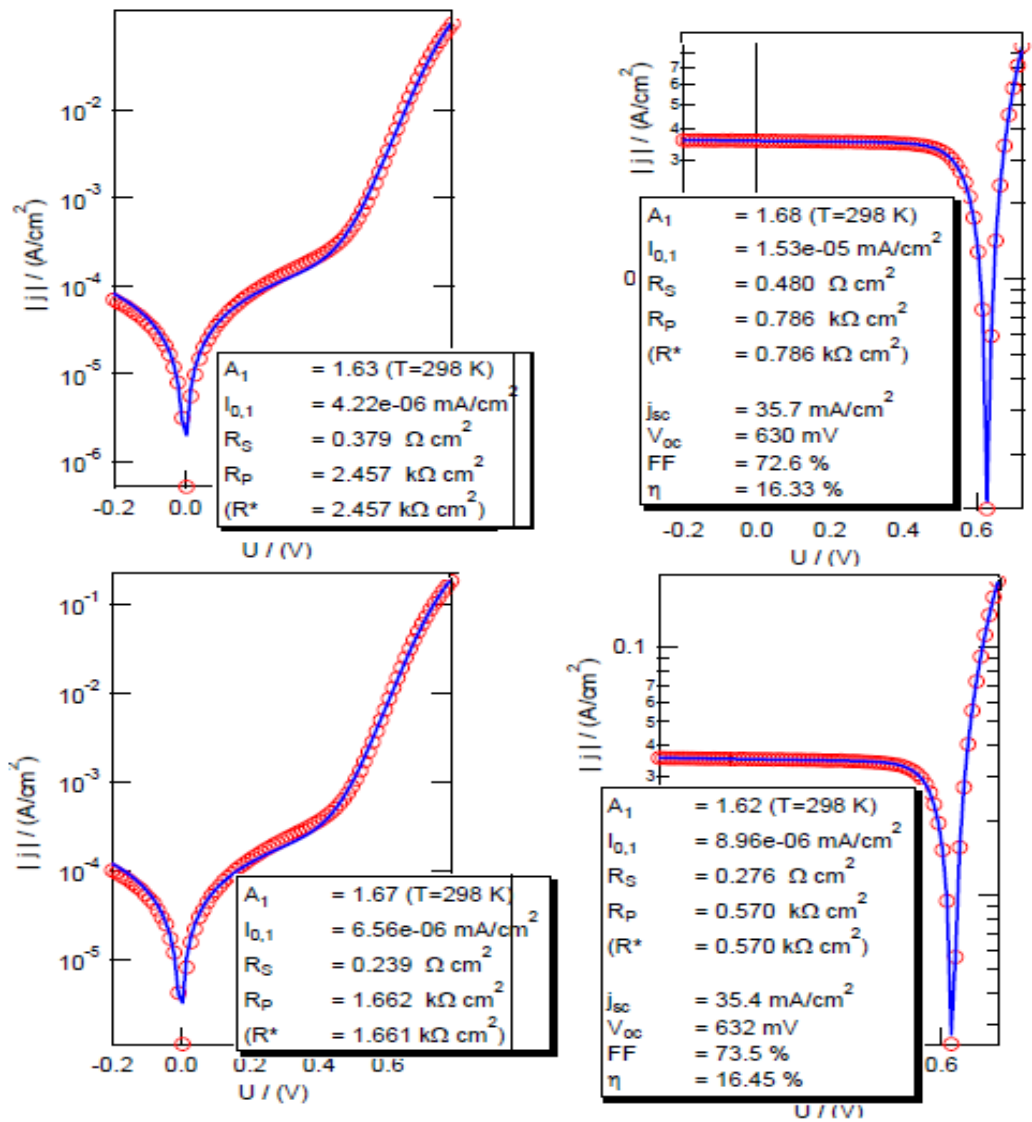

Fig. 7, the current-voltage characteristics of 4 cells measured by the 2-wire probe configuration using the analytical parameter extraction model of Fig. 6 and equation 1.

The lower values extracted for $J_o$ and $n$ further suggests a lesser trap concentration at the interfaces of CIGS or contacts, which in turn, widens the depletion width and fosters the carrier collection (high quantum efficiency). Therefore, the depletion region is dominated by the recombination current and the charge transport in the voltage range of 0.44 V < V < 0.7 V is called recombination current transport. On the other hand, a higher ideality factor of $n_2$=6.3 is extracted for V < 0.45 V which is attributed to recombination current transport outside the depletion region. This is the case for the case n>2 where the quasi-neutral region is dominated by the trap-assisted tunneling and field-assisted recombination. Nevertheless, we believe that in our case, this high ideality factor is due to nonlinear contact resistances which has reduced for 4-wire probe configuration.

Tan et al, suggested different current conduction mechanisms for different voltage regions of device characteristics [17]. For example, lower voltage regions within 0 < V < 0.3 V do not fit the linear straight lines but fit with ¼ powder law given by $I \equiv V^{1.4}$, which is called tunneling current or Space Charge Limited Current (SCLC). The latter mechanism is well defined by Child-Langmuir law [18]. Within mid-voltage regions 0.3V < V < 0.55 V fits with the Shockley-Read-Hall recombination through the bulk region and the distribution of trap states within the depletion width as so-called diode recombination current. This mid-region gives high ideality factors than 1 (n>1) as stated previously [19]. At higher voltages for V > 0.55 V, the SCL current dominates the charge transport. The SCLC occurs due to excessive concentration of electrons passivated into the depletion width which deforms the electric field distribution at the interface [16]. Therefore, the SCLC occurs by the drift component of the carrier conduction current or the injected carriers and is formulated by the Mott-Gurney law. Here, the SCLC follows an exponential distribution of traps outside the depletion width and not at the interface [20]. A 2.43 power-law fits with SCLC in this region $I \equiv V^{2.43}$. The above discussion reminds the importance and significant contribution of $J_0$ and $n$ in the current conduction mechanism and carrier transport. For a low $J_0$ and a low $n$, the recombination current is not strong at low voltage.

In addition to the current conduction mechanism analyzed above, we also investigated if the collection time can influence on series resistant measured by the two probe configurations. We measured the IV curves for a small Δt=0.1 s and a large Δt=10 s. A smaller current density was measured for the small Δt=0.1 s especially at high voltages and that can be attributed to higher series resistance or smaller fill factor [21,22]. This means a higher carrier collection time (ΔT=10 s) can foster the carriers to be collected efficiently by the electrodes (probes) if the voltage is high enough. Either the 4-wire probe can be more efficient as there is more chance for the carriers to pass to the electrode from the cell. In this case, the recombination rate is lower since the photogenerated electrons are rapidly collected into the respective electrode (here as probes), and thus, the current density increases [23,24].

**CONCLUSION**

Thin-film solar cells are now a reliable alternative for the Si photovoltaics. We have studied a CIGS thin-film solar cell using the 2- and 4- wire probe configuration technique using a Cascade probe system under dark and forward bias conditions. We measured the charge collection time or the dead-time (ΔT) from both 2- and 4-wire probe configuration an extracted the series and shunt resistances following an analytical approach presented in the literature for simulation and series or shunt resistance of the cell as well as the ideality factor and saturation current density. Besides, the device metrics were extracted following the analytical theory presented by Das et al. according to a double-diode model. Four parameters were calculated extracted from the measured JV curves at dark: $R_s$, $R_{sh}$, $n$, $J_0$, suggesting a different current conduction mechanism for different applied voltage ranges. A longer Δt means sufficient time to collect the carriers, which gives a slightly higher current density at higher voltages. Lower $J_o$ and $n$ values measured at 300K suggest the recombination current domination representing a quality junction interfaces with a comparatively smaller concentration of surface impurities. The recombination within the space charge region promotes the recombination at the interfaces which in turn help it to govern the carrier transport mechanism within 0.44 V < V < 0.7 V voltage region. We demonstrated the importance and significant contribution of $J_0$ and $n$ in the current conduction mechanism and carrier transport.

**Acknowledgment**

This research is funded by the Foundation for Science and Technology Development of Ton Duc Thang University (FOSTECT), website: http://fostect.tdtu.edu.vn, under Grant FOSTECT.2019.18. We also acknowledge the financial support of this work by the Hungarian State and the European Union under the EFOP-3.6.1-16-2016-00010 project.